\documentclass[final,5p,times,twocolumn]{elsarticle}
\usepackage{amsfonts,bm,amsmath}
\usepackage{graphicx}
\usepackage{color}

\biboptions{sort&compress}

\journal{Physics Letters B}

\usepackage{graphicx}
\usepackage{epsfig}
\usepackage{latexsym}
\usepackage{amsmath}
\usepackage{amsfonts}
\usepackage{amsxtra}
\usepackage{xcolor}

\def\krto{ {\,\,\lower .8ex\hbox {$\longrightarrow \atop k \rightarrow 0$}\,\,}}

\def\bea{\begin{eqnarray} }
\def\beq{\begin{eqnarray} }

\def\eea{\end{eqnarray}}
\def\eeq{\end{eqnarray}}

\def\eq#1{Eq.~(\ref{#1})}

\begin{document}

\title{Gluon Green functions free of Quantum fluctuations}

\author[a]{A. Athenodorou}
\address[a]{Department of Physics, University of Cyprus, POB 20537, 1678 Nicosia, Cyprus}
\author[b]{Ph.~Boucaud} 
\address[b]{Laboratoire de Physique Th\'eorique (UMR8627), CNRS, Univ. Paris-Sud, Universit\'e Paris-Saclay, 91405 Orsay, France}
\author[c]{F.~De Soto}
\address[c]{Dpto. Sistemas F\'isicos, Qu\'imicos y Naturales, 
Univ. Pablo de Olavide, 41013 Sevilla, Spain}
\author[d]{J.~Rodr\'{\i}guez-Quintero}
\address[d]{Dpto. F\'isica Aplicada, Fac. Ciencias Experimentales; 
Universidad de Huelva, 21071 Huelva; Spain.}
\author[e]{S.~Zafeiropoulos}
\address[e]{Institut f\"ur Theoretische Physik, Goethe-Universit\"at Frankfurt,
Max-von-Laue-Str.~1, 60438 Frankfurt am Main, Germany}
\begin{abstract}

This letter reports on how the Wilson flow technique can efficaciously kill the short-distance quantum fluctuations of 2- and 3-gluon Green functions, removes the $\Lambda_{\rm QCD}$ scale and destroys the transition from the confining non-perturbative to the asymptotically-free perturbative sector. After the Wilson flow, the behavior of the Green functions with momenta can be described in terms of the quasi-classical instanton background. The same behavior also occurs, before the Wilson flow, at low-momenta. This last result permits applications as, for instance, the detection of instanton phenomenological properties or a determination of the lattice spacing only from the gauge sector of the theory.  

\end{abstract}

%\pacs{12.38.Aw, 12.38.Lg}

\maketitle

%\begin{flushright}
%%DAMTP-2011-nnn\\
%LPT-Orsay 11-74\\
%UHU-FT/11-29 \\
%%LPSC-11-nnn \\
%IRFU-11-136
%\end{flushright}
%%
%\vspace*{-1cm}
%%
%\begin{figure}[h]
%%  \begin{center}
%    \includegraphics[width=25mm]{figs/ETMC_rund.pdf}
%%  \end{center}
%\end{figure}

%\vfill
%\newpage

%%%%%%%%%%%%%%%%%%%%%%%%%%%%%%%%%%%%%%%%%%%%%%%%%%%%%%%%%
%% end of title page
%%%%%%%%%%%%%%%%%%%%%%%%%%%%%%%%%%%%%%%%%%%%%%%%%%%%%%%%%

\section{Introduction}
%\alinea

%{\it Introduction}.--
QCD, the quantum field theory of strong interactions, is a non-abelian gauge theory with a very rich non-perturbative low-momentum sector where crucial phenomena such as confinement and chiral symmetry breaking take place. An appealing approach to obtain some understanding of this sector is based on describing the gauge fields in terms of short-distance quantum fluctuations on top of topologically non-trivial solutions of the classical field equations in Euclidean space with finite action, the so called instantons~\cite{Belavin:1975fg,'tHooft:1976fv, 'tHooft:1976up}. These solutions shed light into many interesting phenomena such as the explanation of the U(1) problem ~\cite{U(1)}, they can be interpreted as tunneling paths between vacua with different winding number in Minkowski spacetime~\cite{Bitar}, they are related to the strong $CP$ problem ~\cite{Jackiw}, to the lower part of the Dirac operator spectrum and chiral symmetry breaking ~\cite {VS}(for more details we refer the reader to~\cite{Sidney, Shifman, Schafer:1996wv, pvn}). Applications of instantons extend well beyond the scope of QCD such as in the electroweak sector of the Standard Model describing rare processes of baryon decay ~\cite{'tHooft:1976fv}, studies of decays into the true vacuum which could potentially have profound applications in the fate of the early Universe 
%which after the Big Bang at very high temperatures is in the symmetric vacuum state but due to the expansion and cooling the potential could develop the true asymmetric vacuum
 ~\cite{Coleman1, Coleman2}. 
 Instanton applications in supersymmetric theories are also noteworthy, since celebrated results such as the exact $\beta$-function were achieved employing instanton calculus ~\cite{NSVZ}.
%such as $\mathcal{N}=4$ SYM.
Indeed, in most applications, one cannot deal with the exact solutions of the field equations but with approximated quasi-classical field configurations obtained by the minimization of the action.
 %within a given functional space 
These are defined through an ansatz, inspired by the exact one-instanton solution~\cite{Shuryak:1987iz,Shuryak:1987ja,Shuryak:1987tr,Diakonov:1983hh,Verbaarschot:1991sq}.

In practice, these quasi-classical field configurations have been ``observed" by means of numerical simulations in 
lattice QCD~\cite{Trewartha:2013qga,deForcrand:1997ut,GarciaPerez:1993lic,deForcrand:1997esx,Negele:1999mb}. In addition, local recognition of instantons' geometrical shapes around their centers, after applying a cooling procedure perceived to eliminate quantum fluctuations~\cite{Teper:1985rb}, has also been addressed extensively. Cooling is a discrete method based on making successive "sweeps" to the lattice configuration of fields, known to minimize the action but also to introduce biases that could potentially lead to uncontrollable effects. Although a number of different alternatives has been proposed to prevent these effects (e.g. \cite{Smith:1998wt,Negele:1998ev}), the so-called Wilson flow has been recently proposed as a theoretically well founded smoothing technique~\cite{Luscher:2010iy} that encompasses many attractive features with the main one being that the “flown” fields renormalize in a very simple fashion ~\cite{Luscher:2011bx}. 

On the other hand, a few lattice studies focused on the identification of the effects originating from the quasi-classical instanton contribution on gluon correlation functions and to investigate whether such effects can be potentially distinguished, before applying any smoothing technique, within a given low-momentum window~\cite{Boucaud:2002fx,Boucaud:2003xi,Boucaud:2004zr}. In doing the latter, avoiding the smoothing procedure as it might distort the gauge fields, two main goals can be achieved. One can advocate strongly in favor of the presence of quasi-classical structures (and even their low-momentum dominance) in gauge configurations. Moreover, some instantonic properties without the need of any sort of extrapolation to the physical non-smoothed situation can be measured. 

In the present letter, we will compute and analyze the two- and three-point gluon Green functions in momentum space. The results obtained before and after applying the Wilson flow will be compared, with the first main objective to unravel the behavior as a function of the momentum, in the whole momentum range, that survives the annihilation of short-distance fluctuations. Next, we pinpoint whether the same behavior dominates the gluon correlations at low momenta, and finally we aim at an interpretation in terms of instantons. In order to achieve the last on the basis of the most general assumptions, we focus on the study of a particular combination of two- and three-point Green functions defining a three-gluon running coupling in the momentum subtraction (MOM) scheme. Furthermore, all the past lattice studies of gluon correlations in terms of instantons had been made in the quenched approximation, i.e.\ without dynamical quarks. Here, gauge fields obtained from both quenched and unquenched lattice simulations will be analyzed and compared.

%-------------------------------------------------------
%-------------------------------------------------------

\section{Wilson flow}
\label{sec:WF}

%{\it Wilson flow}.--
Let us start by a very brief introduction to the Wilson flow, which has proven to be an essential tool in modern non-perturbative studies of QCD~\cite{Luscher:2010iy,Luscher:2013vga}. It is easier to analyze it first in continuum language, before introducing its lattice counterpart. 

Like many other techniques that have been developed in the past decades in order to efficiently deal with unphysical short-distance fluctuations, also the Wilson flow can be conceived as a smoothing procedure which diminishes these unphysical fluctuations.
% and which can isolate the physical content of the theory. 
However, in the framework of a quantum field theory, short-distance corresponds to ultra-violet (UV) quantum fluctuations and depriving the gauge fields from them, potentially, implies to isolate the underlying non-trivial classical solutions which minimize the gauge action. 

The Wilson flow $B_{\mu}(t,x)$ of an SU(N) gauge field is defined by the following first order differential equation~\cite{Luscher:2009eq,Luscher:2010iy,Luscher:2013vga}
%-----
\begin{eqnarray}
\partial_{\tau} B_{\mu}=D_{\nu}G_{\nu\mu},
\end{eqnarray}
%-----
where $\tau$ is the so-called flow time and
%-----
\begin{eqnarray}
\hspace{-0.5em}G_{\mu\nu}&=&\partial_{\mu}B_{\nu}-\partial_{\nu}B_{\mu}
+[B_{\mu},B_{\nu}], \\
\hspace{-0.5em} D_{\mu}&=&\partial_{\mu}+[B_{\mu},\,\cdot\,],
\end{eqnarray}
with  the initial condition $B_{\mu}(0,x)=A_{\mu}(x)$. The expansion of the flown field $B_{\mu}(\tau,x)$ in terms of the fundamental field $A_{\mu}(x)$ reads 
%-----
\begin{eqnarray}
B_{\mu}(\tau,x) &=& \int d^4y~K(\tau;x-y) A_{\mu}(x)\,, \\ 
K(\tau;x) &=& \frac{e^{-\frac{x^2}{4\tau}}}{\left(4\pi \tau\right)^{2}}\,,
\end{eqnarray}
where smoothing is destroying short-distance fluctuations (at tree-level) over a radius of $\sqrt{8\tau}$. 

The lattice counterpart of the Wilson flow, previously introduced in the context of Morse theory~\cite{Atiyah:1982fa}, is defined (see \cite{Luscher:2009eq,Luscher:2010iy}) by the solution of the differential  equation
  \begin{equation}
  \begin{aligned}
  &\partial_\tau V_{\mu}(x,\tau)=- g_0^2\big[\partial_{x,\mu}S(V(\tau))\big]V_{\mu}(x,\tau) \\
  & V_{\mu}(x,0)=U_{\mu}(x)\ ,
 \end{aligned}
\end{equation}
%-------
where $S$ is some discretization of the gauge action
%, $\tau$ is the lattice flow time
 and $g_0$ the bare coupling. A definition of the link derivatives $\partial_{x,\mu}$ can be found in ref. ~\cite{Luscher:2010iy}. From a historic viewpoint the “streamline” idea of Refs.~\cite{Shuryak:1987iz,Shuryak:1987ja,Shuryak:1987tr} is intimately related to the idea of the gradient flow.

Besides the important features of existence, uniqueness and smoothness of the flow~\cite{Luscher:2010iy} another very attractive feature of the flow is the fact that expectation values of local observables built out of the “flown” fields assume a well defined continuum limit. It is important to mention that, in order to avoid composite operators’ renormalization, 
those observables should be evaluated at fixed flow time in physical units while taking the continuum limit.

Other smoothing techniques such as the usual cooling or continuous version of smearing had been previously proposed~\cite{Narayanan:2006rf} and, very recently, a perturbative equivalence between flow time and number of cooling steps has been established through the comparison of the topological charge obtained with both cooling and Wilson flow~\cite{Bonati:2014tqa, Alexandrou:2015yba}.

\section{Lattice Green functions}
\label{sec:Lat}

%{\it Lattice Green functions}.--
Now, as explained in ref.~\cite{Boucaud:1998bq}, we will compute from lattice QCD simulations the MOM three-gluon coupling defined as
%----------
\begin{equation}\label{eq:3g0}
\alpha^{\rm 3-g}(k^2) = \frac{k^6}{4\pi} \frac{\left( G^{(3)}(k^2)\right)^2}{\left(G^{(2)}(k^2)\right)^3} 
 \ ,
\end{equation}
%----------------
where
%-----------
\begin{equation}
G^{(m)}(k^2) = \frac 1 N T^{\mu_1,\dots \mu_m}_{a_1,\dots a_m} \langle 
\widetilde{A}_{\mu_1}^{a_1}(k_1) \dots 
\widetilde{A}_{\mu_m}^{a_m}(k_m)
\rangle \  
\end{equation}
%------------
stands for the $m$-point Green function in Landau gauge, $\widetilde{A}_\mu^a$ is the gauge field in momentum space, $a$ ($\mu$) are color (Lorentz) indices and $T$ and $N$ are the tree-level tensor and normalization factor needed for the appropriate projection in each case (for instance $T^{\mu_1,\mu_2}_{a_1,a_2} = \delta_{a_1,a_2} \left( \delta^{\mu_1 \mu_2}- k^{\mu_1} k^{\mu_2}/k^2\right)$ and $N$=24 for $m=2$). The kinematical configuration for the Green functions is chosen to satisfy: $\sum_i^m k_i= 0$ and $k_i^2=k^2$ $\forall$ $i=1,\dots m$. 

Then, we can obtain the gauge fields directly from an ensemble of lattice configurations, as done in ref.~\cite{Boucaud:1998bq}, compute the correlation functions and the coupling defined by \eq{eq:3g0}. A main advantage of analyzing this particular coupling is that it offers the renormalization group invariant (RGI) combination of two- and three-point Green function from the RHS of \eq{eq:3g0}, which keeps no dependence on either the regularization parameter (as is implicitly the case for the m-point lattice Green functions) or the renormalization momentum, if any renormalization prescription is applied. The gauge fields can be obtained before or after the Wilson flow for any flow time. At any step, before and after applying the Wilson flow, in order to get the gauge-fixed Green functions that should be plugged into \eq{eq:3g0}, the gauge fields should be properly brought to the Landau gauge. 

In our results, we have exploited unquenched lattice configurations with two degenerate light dynamical flavors ($u$ and $d$) and two heavier ($s$ and $c$) flavors which made possible a successful determination of the $\overline{\rm MS}$ running coupling at the $Z^0$-mass scale~\cite{Blossier:2012ef}. We have obtained new quenched configurations at several large volumes and different bare couplings. 600 configurations at $\beta=3.90$ for a 64$^4$ lattice volume (15.6$^4$ fm$^4$) and 220 at $\beta=4.20$ for 32$^4$ (4.5$^4$ fm$^4$), all of them employing the tree-level Symanzik gauge action; and 380 at $\beta=2.37$ for 20$^3\times$40 (2.8$^3\times$ 5.6 fm$^4$), with the Iwasaki gauge action. The idea behind using different gauge actions relies to the clarification that the a priori different cut-off effects should not pose any concern. In the unquenched case, we have used 200 configurations at $\beta=1.95$ for a 48$^3\times$96 lattice volume (4.0$^3\times$7.9 fm$^4$), a pion mass of $297$ MeV with the Iwasaki gauge action and the Twisted Mass action in the fermionic sector. More details for the set-up and specifics of the unquenched configurations, can be found in~\cite{etmc, Ayala:2012pb}.

\section{Multi-instanton background}
\label{Inter}

%{\it Multi-instanton background}.--
We will analyze the results in terms of the quasi-classical solutions of the SU(3) gauge action. In ref.~\cite{Shuryak:1987iz}, the gauge-field classical solution from an ensemble of instantons, $B_\mu^a$, has been proposed to be cast as the following trial function, 
%------------
\beq\label{r-ansatz}
g_0 B_\mu^a(x) \ = \ \frac {\displaystyle 2 \sum_{i=I,A} R_{(i)}^{a\alpha} \overline{\eta}^\alpha_{\mu\nu} 
\frac {y_i^\nu}{y_i^2} \ \rho_i^2 \frac{f(|y_i|)}{y_i^2} } 
{\displaystyle 1 + \sum_{i=I,A} \rho_i^2 \frac{f(|y_i|)}{y_i^2}} \ ,
\eeq
%------------
coined as the ratio-ansatz, where $y_i=(x-z^i)$ and $\overline{\eta}^\alpha_{\mu\nu}$ is the 't~Hooft symbol, that should be replaced by $\eta^\alpha_{\mu\nu}$ when summing over anti-instantons as $i=A$. $R_{(i)}^{a\alpha}$ represents the color rotations embedding the canonical SU(2) instanton solution in the SU(3) gauge group ({\it i.e., $\alpha=1,2,3$} and $a=1,2, \dots 8$). 
$f(x)$ is a shape function that obeys $f(0)=1$ in order not to spoil the 
field topology at the instanton centers which also provides sufficient cut-off at large distances guaranteeing convergence of the sum. 

Two particular asymptotic limits can be identified in \eq{r-ansatz}. First,  if the gauge field is evaluated far away from all instantons' centers, {\it i.e.} for any $x$ such that $y_i >> \rho_i$ for all $i$, the aforementioned large-distances cut-off makes the shape function to drop off keeping only the unity in the denominator and one is left with
%----------------
\beq
g_0 B_\mu^a(x) \ \sim \ 2 \sum_{i=I,A} R_{(i)}^{a\alpha} \overline{\eta}^\alpha_{\mu\nu} 
\frac {y_i^\nu}{y_i^2} \  \rho_i^2 \frac{f(|y_i|)}{y_i^2} \ .
\eeq
%------------------
On the other hand, as the gauge field is evaluated near one given instanton or anti-instanton labeled 
with $i=j$, {\it i.e.} for any $x$ such that 
$y_j << \rho_j$, while $y_i >> \rho_i$ for any $i \neq j$, 
\bea\label{xlow}
g_0 B_\mu^a(x) &\sim& 2  R_{(j)}^{a\alpha} \overline{\eta}^\alpha_{\mu\nu} 
\frac {y_j^\nu}{y_j^2} \ \frac 1 {1+\frac {y_j^2}{\rho_j^2}} \nonumber \\
&\sim&
2  \sum_{i=I,A} R_{(i)}^{a\alpha} \overline{\eta}^\alpha_{\mu\nu} 
\frac {y_i^\nu}{y_i^2} \  \frac {f(|y_i|)} {f(|y_i|) +\frac {y_i^2}{\rho_i^2}}  \ .
\eea
%-------------------
Thus, in both the regimes of large and small distances, the gauge 
field  can be effectively described by the following independent-pseudoparticle sum-ansatz approach, 
%-----------------
\begin{equation}
g_0\, B_{\mu}^{a}({\bf x})= 2 \sum_i R^{ a \alpha}_{(i)}\,
\overline{\eta}^\alpha_{\mu \nu} \ \frac {y_i^\nu}{y_i^2} \
\phi_{\rho_i}\left(\frac{|y_i|}{\rho_i}\right) \ , \label{amuins}
\end{equation} 
%----------------
provided that the profile function $\phi$ behaves as
%----------------
\bea\label{eq:match}
\phi_\rho(z) = \left\{ 
\begin{array}{lr} 
\displaystyle \frac{f(\rho z)}{f(\rho z)+z^2} \simeq \frac 1 {1+z^2} & z \ll 1
 \\
\displaystyle \frac{f(\rho z)}{z^2} & z \gg 1
\end{array} \right. \ ,
\eea
%----------------
where $f(z)$ is the shape function which can be obtained by minimizing the action per particle for some statistical ensemble of instantons defining the semi-classical background. This function essentially drives the large-distance behavior of the gauge field due to one-instanton contributions and incorporates also the nonlinear effects resulting from the average classical interaction of the other instantons in the background. According to \cite{Diakonov:1983hh}, this shape function and the large-distance drop can be approximated as being independent of the low-distance scale $\rho$ fixing the instanton size. However, the profile function $\phi$, defined to match both large- and low-distance behaviors, needs to break this scale independence as we did explicitly in \eq{eq:match}.

Then, as explained in \cite{Boucaud:2002fx}, the gauge-field Green functions can be semi-classically obtained within the instanton background as
%-----------
\begin{equation}\label{eq:Gm}
g_0^m G^{(m)}(k^2) =  \frac{k^{2-m}}{m 4^{m-1}} \ n \ \langle \ \rho^{3m} I^m(k\rho) \ \rangle 
\end{equation}
%------------
where  $n$ is the instanton density,  
\begin{equation}
I(s) \ = \ \frac{8 \pi^2}{s} \int_0^\infty z dz J_2(s z) \phi(z) \ ,
\end{equation}
and $\langle \dots \rangle$ expresses the average over the distribution of instantons within the statistical ensemble defining the background. 

Thus, one would have
%---------------
\begin{equation}\label{eq:3g}
\alpha^{\rm 3-g}(k^2) = \frac{k^6}{4\pi} \frac{\left( G^{(3)}(k^2)\right)^2}{\left(G^{(2)}(k^2)\right)^3} 
= \frac{k^4}{18 \pi n} \frac{\langle \rho^9 I^3(k\rho)\rangle^2}{\langle \rho^6 I^2(k\rho)\rangle^3} \ .
\end{equation}
%-----------------
Whichever the shape function $f(x)$ might be, the topological condition $f(0)=1$ guarantees that $I(s)$=$18\pi^2/s^3$ when $s\to \infty$ and then 
%---------
\begin{equation}\label{eq:large}
\frac{\langle \rho^9 I^3(k\rho)\rangle^2}{\langle \rho^6 I^2(k\rho)\rangle^3} \simeq 1 + 
{\cal O}\left(\frac{\delta\rho^2}{k^2\bar{\rho}^4} \right) \ ,
\end{equation}
where $\bar{\rho}=\sqrt{\langle \rho^2 \rangle}$ and $\delta\rho^2=\langle(\rho-\bar{\rho})^2\rangle$ stand for the mean square width of the radii distribution. On the other hand, only relying on the sufficient cut-off of $f(x)$ at large distances, one would be left with
%---------------
\begin{equation}\label{eq:dist}
\frac{\langle \rho^9 I^3(k\rho)\rangle^2}{\langle \rho^6 I^2(k\rho)\rangle^3} \simeq 
1+48 \frac{\delta\rho^2}{\bar{\rho}^2} + {\cal O}\left(k^2\delta\rho^2,\frac{\delta\rho^4}{\bar{\rho}^4}\right) \ .
\end{equation}
for the low-momentum domain. Notice that, had we considered a zero width for the radii distribution, the coupling defined by \eq{eq:3g0} would plainly behave as a scale-independent $k^4$-power law for all momenta. 

\section{Results and discussion}

\begin{figure}[h]
\begin{center}
\includegraphics[width=0.9\linewidth]{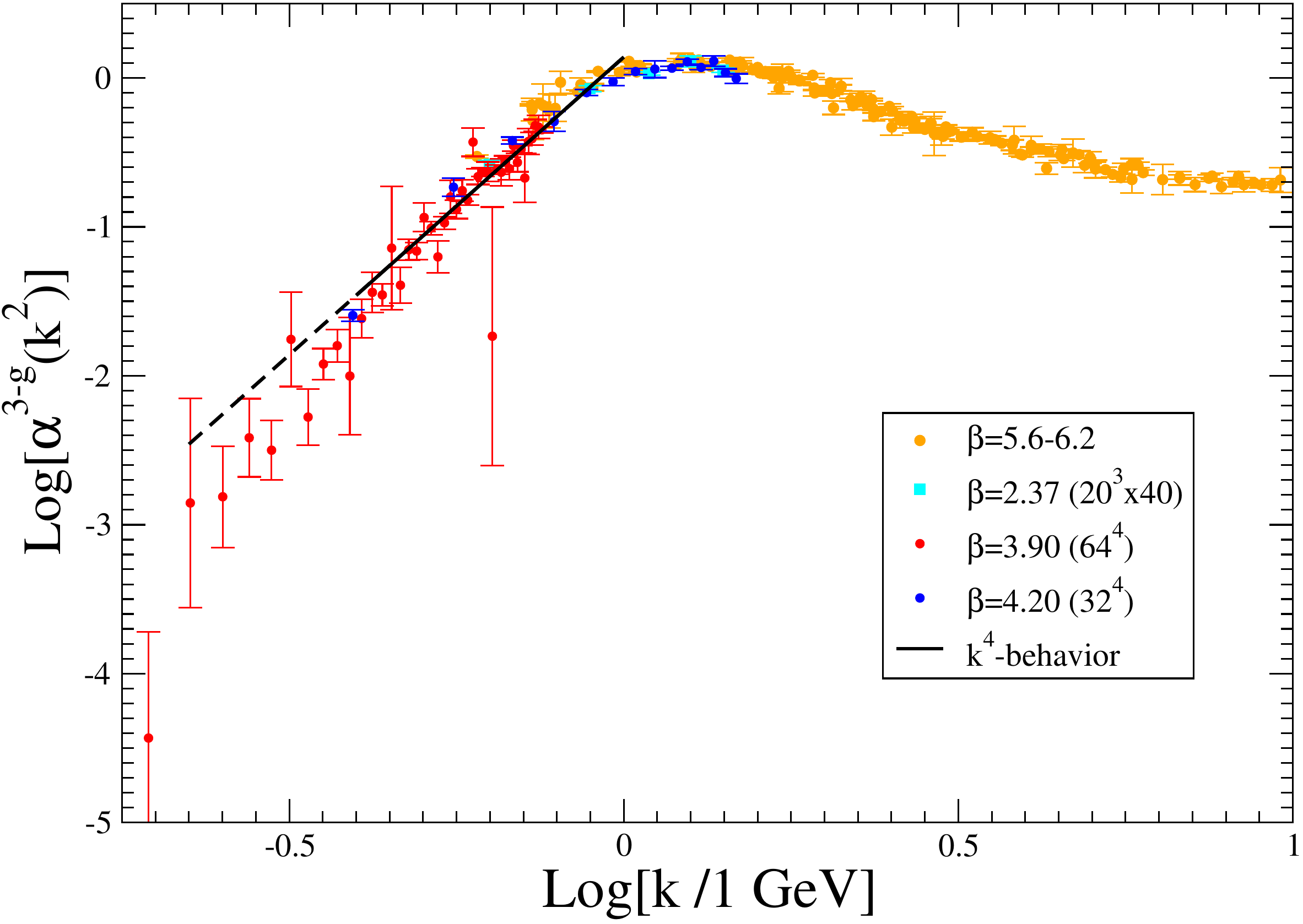} 
%\vspace*{-0.25cm}
\caption{\small The MOM three-gluon coupling defined in \eq{eq:3g0} obtained from all the different quenched lattice simulations described in the text. } 
\label{fig:all}
\end{center}
\end{figure}

%{\it Results and discussion}.--
The coupling obtained according to \eq{eq:3g0} for all quenched simulations at zero flow time appears displayed In Fig.~\ref{fig:all}. On top of it, for the sake of comparison, we have also incorporated additional data (orange solid circles) for the same coupling obtained from simulations in much smaller lattice volumes (ranging from 2.4$^4$ to 5.9$^4$ fm$^4$) with the Wilson gauge action for several $\beta$'s from 5.6 to 6.0, and published more than a decade ago~\cite{Boucaud:2002fx}.  It should be first noticed that, as corresponding to the RGI nature of the RHS of \eq{eq:3g0}, all the data from different simulations with different actions and set-up's show a very good physical scaling. However, the main feature to be underlined is that, before applying the Wilson flow, a momentum scale, lying around 1 GeV (in the ballpark of $\Lambda_{\rm QCD}$), separates clearly two regimes, the one above this scale where quantum corrections manage to build the well-known perturbative logarithmic running and that below, where the power law from \eq{eq:3g} appears to rise. The intercept of the low-momenta logarithmic line, as is highlighted by the above-mentioned good scaling, is a physical quantity, and can be very well used for a cheap calibration of the lattice spacing. Its value estimated from data is 1.44 GeV$^{-4}$ and, by neglecting the radii distribution width, one would be left for the instanton density with $n=7.7(1)$ fm$^{-4}$.

%\vspace*{-0.5cm}

\begin{figure}[h]
\begin{center}
\begin{tabular}{c}
\includegraphics[width=0.9\linewidth]{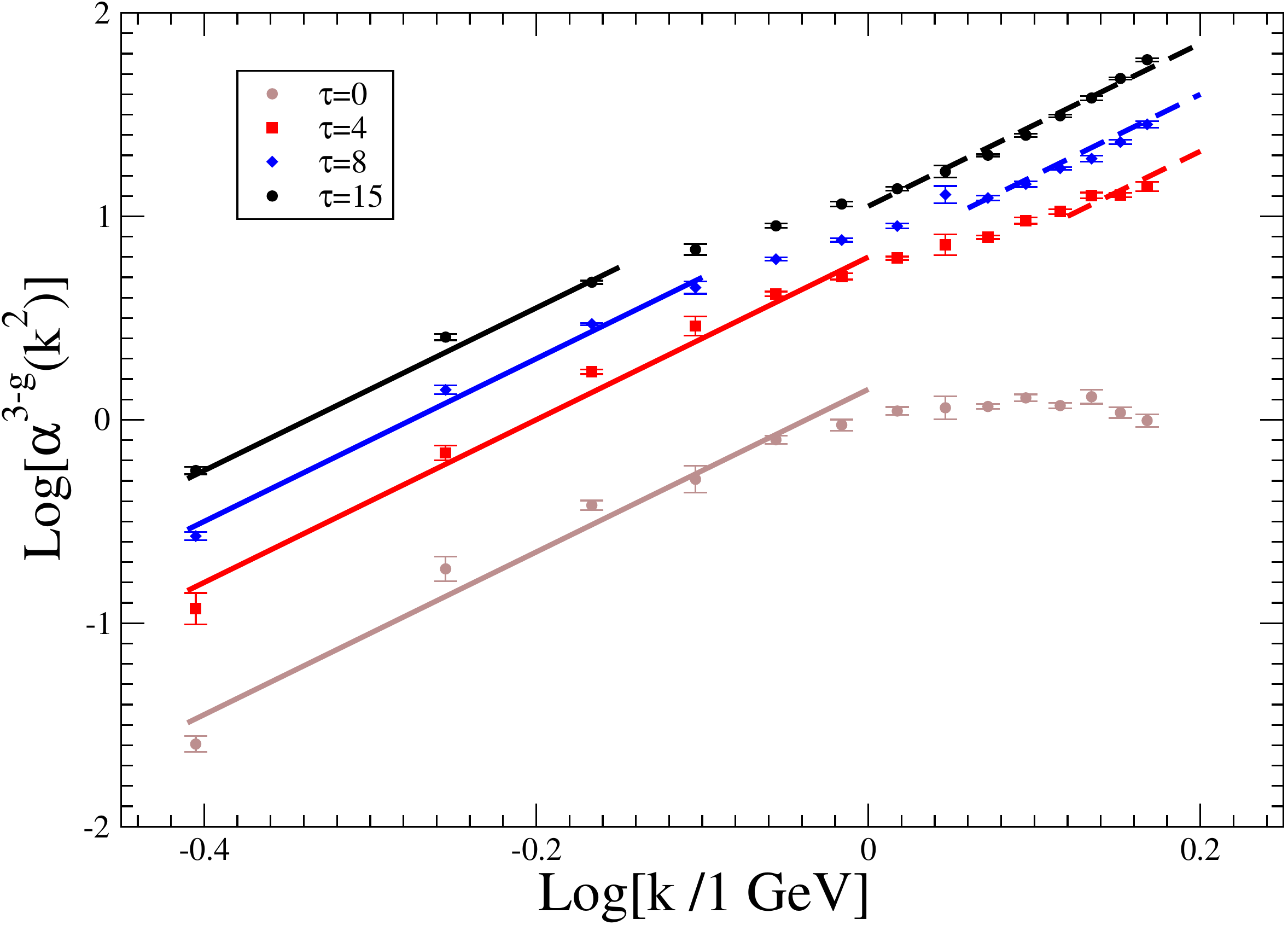} 
\\ 
\includegraphics[width=8cm]{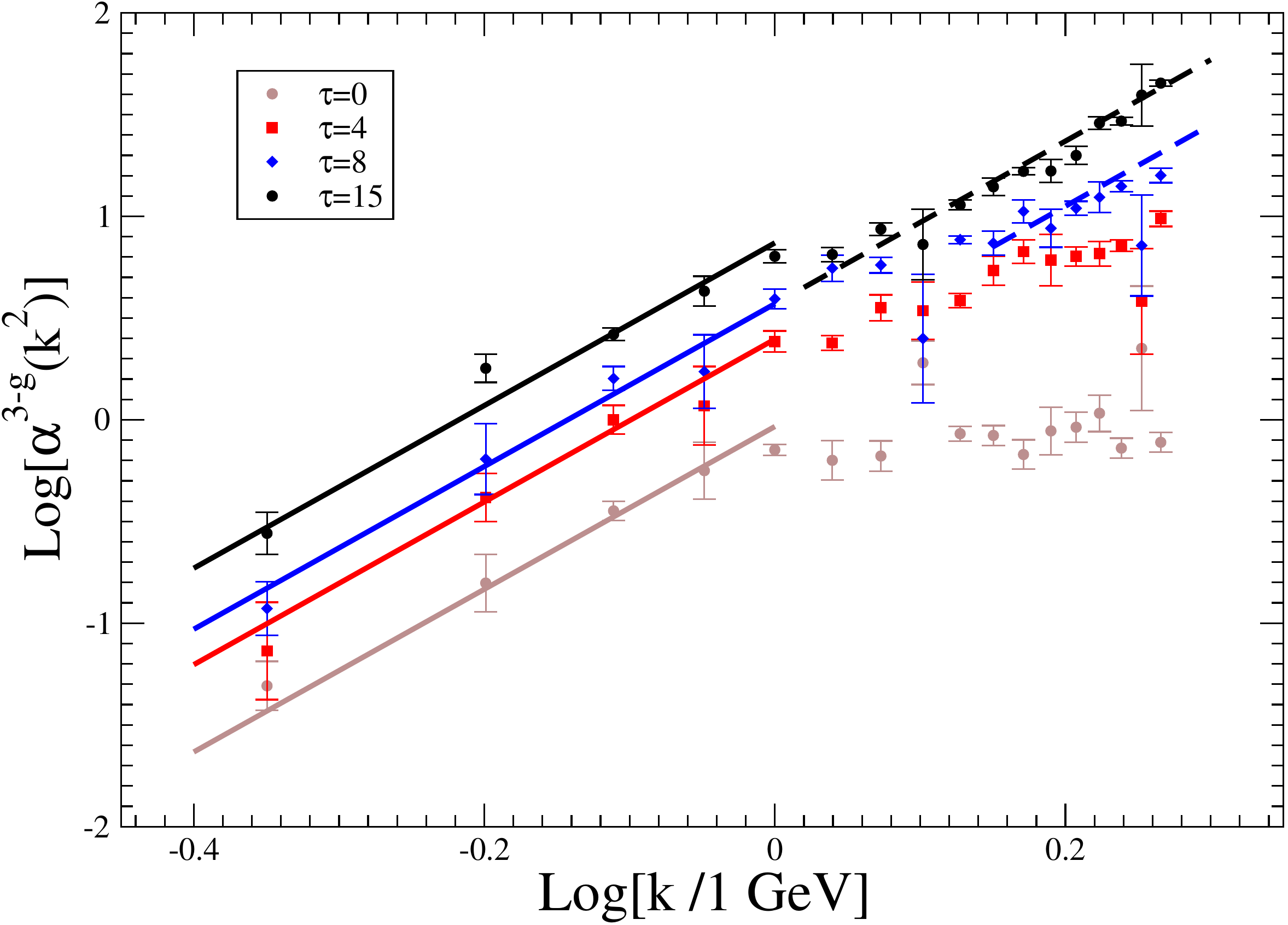} 
\end{tabular}
%\vspace*{-0.25cm}
\caption{\small The MOM three-gluon coupling defined in \eq{eq:3g0} obtained from quenched data with $\beta=4.20$ (top panel) and unquenched with $\beta=1.95$ (bottom panel) lattice simulations at different flow times.} 
\label{fig:regime}
\end{center}
\end{figure}

We have then applied the Wilson flow, for three different flow times ($\tau$=4,8 and 15), to the quenched lattice configurations at $\beta=4.20$ and the unquenched ones at $\beta=1.95$, computed the coupling and displayed the results in Fig.~\ref{fig:regime}.  There, Eqs.~(\ref{eq:3g}-\ref{eq:dist}) explain the $k^4$-behavior observed in both the low- and large-momentum domains. The intercepts of the large-momentum lines provide with an estimate for the instanton density at different flow times: $n=3.5(1),1.75(4)$,$0.98(5)$ fm$^{-4}$ at $\tau=4,8$,$15$, for the quenched case; and $n=6.8(5),3.0(2)$ fm$^{-4}$ at $\tau=8,15$, for the unquenched case (see Tab.~\ref{Tab:densities}, where the flow time is also approximately expressed in physical units). Furthermore, the larger the flow time the lower momenta the non-enhanced linear behavior of \eq{eq:large} appears to extend down for. This suggests that the instanton size grows with the flow time, at least in a first stage, when the instanton density is as high as we obtain and the instanton-anti-instanton annihilation is the mechanism dominating the evolution of the quasi-classical solutions. In order to confirm the estimates of instanton densities here obtained, independent shape-dependent direct and indirect methods can also be used. Furthermore, after the successful description of the RGI combination of two- and three-points Green functions defining a coupling with \eq{eq:3g}, one can also apply \eq{eq:Gm} to account separately for each. Although, to this purpose, one would also need to get or model the shape function. In doing this, as the instanton density has been already fixed by the coupling analysis, the only additional free parameter is the instanton size, which would be then obtained from the gauge-sector Green functions without the need of applying any smoothing procedure. This is however the object of a further work~\cite{newfromus}, as we only focus here on the most general shape-independent results.

\begin{table}[ht]
\begin{center}
\begin{tabular}{|c|c|c|c|}
\hline
& $\tau$ & $t/t_0$ & $n$ fm$^{-4}$ \\
\hline
quenched & 4 & 6.84 & 3.5(1) \\
& 8 & 13.7 & 1.75(4) \\
& 15 & 25.6 & 0.98(5) \\
\hline
unquenched & 4 & 2.34 & - \\
& 8 & 4.70 & 6.8(5) \\ 
& 15 & 8.84 & 3.0(2) \\
\hline 
\end{tabular}
\end{center}
\caption{\small Estimates for the densities, obtained as explained in the text, for the different flow times, also expressed in physical units. For this to be done, according to \cite{Luscher:2010iy}, we have defined $\sqrt{8 t_0}=0.3$ fm, whence $t_0=a^2 \tau_0=0.0113$ fm$^2$ and $t=\frac \tau {\tau_0} t_0$. At $\tau$=4, in the unquenched case, the characteristic diffusion length is so small that quantum fluctuations have not been properly removed yet.}
\label{Tab:densities}
\end{table}

On the other hand, according to \eq{eq:dist}, wherever the momenta satisfy $k^2\delta\rho^2 \ll 1$, the intercept for the low-momentum line is shifted up by $\log(1+48\delta\rho^2/\bar{\rho}^2)\simeq 48/\ln10 \ \delta\rho^2/\bar{\rho}^2$. 
Therefore, one can get $\delta\rho^2/\bar{\rho}^2 \simeq 0.014$ (quenched) and $0.013$ (unquenched), from the comparison of the intercepts in Fig.~\ref{fig:regime}. These numbers can be compared to those estimated in~\cite{Smith:1998wt}, by applying direct instanton detection after cooling lattice gauge configurations obtained in the quenched approximation. Therein, in Tab.~6 and 7, the distribution half width, $\sigma_p$, for several lattice set-up's is given. For instance, at $\beta=6.2$,  $\sigma_p/\bar{\rho}$ is found to range from 0.18 to 0.22, for different number of cooling steps; and at $\beta=6.4$ the results range from 0.17 to 0.21. The knowledge of the full distribution is required for a precise conversion of the mean into half width. In literature, investigations of the instanton size distribution can be found where both semiclassical and lattice approaches have been followed (see, for instance, \cite{Diakonov:1983hh,Schafer:1996wv,Verbaarschot:1991sq}). In particular, the authors of ref.~\cite{Ringwald:1999ze} made a careful quantitative analysis where the size distribution is shown to agree well with a two-loop RG improved prediction from instanton perturbation theory. For our purposes here, a rough estimate is however enough and can be made by assuming a gaussian distribution: $\sigma_p/\bar{\rho}=\sqrt{2 \ln{2} \ \delta^2\rho/\bar{\rho}^2} \simeq 0.14$, lying well in the right ballpark.  

Finally, at zero flow time, the unquenched instanton density can be estimated to be 1.55 times larger than the quenched one, if both unknown distribution widths are taken to be the same, from the difference between the intercepts. This number however relies on how sensible is the quenched lattice calibration. 

Thus, studying an RGI combination of Green functions as that in \eqref{eq:3g0} defining the three-gluon coupling leads to strong conclusions about the effects of the the multi-instanton background, as they can be obtained on the basis of very general results, particularly not affected by the shape function for the pseudo-intanton solution. Nevertheless, the drawback is that it can only give access to the instanton density and size distribution width and their variations with the flow time. Other properties related to the semiclassical background, as the instanton size or its full distribution require other approaches for their determination, out of the scope of this paper. 

%\vspace*{-0.5cm}

\section{Conclusions}

%{\it Conclusion}.--
In summary, the results presented here, relying on a very general and firm ground, strongly support that the classical solutions of the SU(3) gauge theory explain the pattern exhibited by two- and three-gluon Green functions either at low-momenta or, after the efficient killing of the UV fluctuations around the classical minima of the theory, for all momenta. The removal of UV fluctuations by the Wilson flow gets rid of the fundamental QCD scale, $\Lambda_{\rm QCD}$, introduced at the quantization level of the theory. The only remaining scale is then the instanton size, $\bar{\rho}$, still fixed by the lattice scale setting,  done before the removal. Whichever mechanism driving the transition from the asymptotically-free large-momentum to the confined low-momentum domain is also removed. 

The dominance of the instanton background on the low-momentum gluon correlations opens the door to some applications as the determination of the instanton density or, after modelling the shape function, the instanton size. A determination of the lattice spacing, anchored only to the gauge sector of the theory, is also possible.

\section*{Acknowledgements} 

We thank the support of Spanish MINECO FPA2014-53631-C2-2-P  research project, SZ acknowledges support by the Alexander von Humboldt foundation. We thank K. Cichy, M. Creutz, O. P\`ene, O. Philipsen, M. Teper, J. Verbaarschot for fruitful  discussions. Numerical computations have used resources of CINES and GENCI-IDRIS
 %under the allocation 2016-052271
as well as resources at the IN2P3 computing facility in Lyon. We are grateful to the European Twisted Mass collaboration for making their gauge field configurations publicly available.

%%%%%%%%%%%%%%%%%%%%%%%%%%%%%%%%%%%%%%%%%%%%%%%%%%%%%%%%%
%%                              BIBLIO
%%%%%%%%%%%%%%%%%%%%%%%%%%%%%%%%%%%%%%%%%%%%%%%%%%%%%%%%%

\section*{References}

\bibliographystyle{ieeetr}
%\bibliography{total}

\end{document}